\documentclass{camera}
\usepackage{graphicx}

\begin{document}

%
\title{Isospin effects in the thermodynamics of finite nuclei}

%
\author{Gr\'egory Lehaut ; Francesca Gulminelli \and Olivier Lopez}

%
\organization{LPC Caen, ENSICAEN, Universit\'e de Caen, CNRS/IN2P3, Caen, France}

\maketitle

\begin{abstract}

It has been proposed that multifragmentation can be related to the liquid-gas 
phase transition of nuclear matter.
We study the statistical properties of finite nuclear matter near the phase 
transition with the help of a Lattice Gas Model (LGM).
The original version of LGM with only one type of charge-neutral particles is 
well known to feature the properties of the liquid-gas phase transition. 
In this contribution, we address the effect of Coulomb and isospin dependence
 interaction for the finite nuclei transition, and study the symmetry energy 
properties of finite temperature systems.\\

\end{abstract}

%

In the last decade, there has been a growing interest for the measurement of the symmetry coefficient $c_{sym}$ of the nuclear equation of state, through experiments with heavy ion collisions.

At zero temperature, most mean-field-based nuclear equations of state can be approximately reproduced through the density ($\rho$) functional form \cite{Chen} :

\begin{equation}
  c_{sym}(\rho) = c_{sym}^0 \left( \frac{\rho}{\rho_0} \right)^\gamma
\label{equCsymRho}
\end{equation}

where $\rho_0$ is the saturation density, and
$\gamma$ is a coefficient which determines the stiffness of the symmetry 
coefficient.
The values of $\gamma$ can widely differ among different EOS parametrisations ($0.5 \div 2$) 
and are not presently constraint by experimental measurements.

One proposed technique is the so-called isoscaling analysis.
It is well known \cite{Tsang} that the ratio $R_{21}(N,Z)$ of isotope 
yields $Y_i(N,Z)$ measured in two reactions similar in temperature but 
different in isospin labelled ($1$,$2$) has an exponential behavior according 
to :

\begin{equation}
R_{21} (N,Z) = \frac{Y_2(N,Z)}{Y_1(N,Z)}\propto exp \left( \alpha N + \beta Z \right)
\label{equIsoscaling}
\end{equation}

where $\alpha$ and $\beta$ are the isoscaling parameters.

Different data have been published showing that $\alpha$ decreases with 
increasing the violence of the collision~\cite{Tsang,Shetty,Lefevre}. 
Since an increasing collision violence is associated to increasing temperature
and/or decreasing density at the time of fragment formation, this suggests that we may access the density and/or temperature evolution of 
the symmetry term.
Indeed at thermal equilibration and in the framework of the grand canonical ensemble, $\alpha$ and $c_{sym}$ are approximately related by \cite{Botvina} : 

\begin{equation}
\alpha T \approx 4 c_{sym} \left( \left(\frac{Z_1}{A_1}\right)^2 - \left(\frac{Z_2}{A_2}\right)^2 \right)
\label{equBotvina}
\end{equation}

where $T$ is the common temperature of the two systems, and $Z_i$,$A_i$ are the 
corresponding charges and masses.

Heavy ion collisions are probably the best probe of nuclear properties in 
excitation energy and density conditions far from the ground state.
However, the connexion between the symmetry energy coefficient appearing in 
eqs.\ref{equCsymRho} and \ref{equBotvina} is far from being clear.
Indeed the fragmented configurations accessed by isoscaling analysis are 
dishomogeneous and present high order correlations; it is therefore not clear 
whether, even at thermal equilibrium, the associated energy functional only 
depends on the global density and temperature as obtained in a mean-field based 
picture.
Moreover eq.\ref{equBotvina} has been derived in the framework of
 macroscopic statistical models~\cite{Botvina} where many body correlations are
 supposed to
be entirely exhausted by clusterisation and clusters are described as 
independent degrees of freedom.
If the validity of isoscaling eq.\ref{equIsoscaling} in microscopic theories
is well settled~\cite{Colonna}, the connexion of $\alpha$ and $c_{sym}$ eq.\ref{equBotvina}, in such models has never been proved.

To progress on these issues, it is interesting to consider a microscopic model 
simple enough to be exactly solvable through Monte-Carlo simulations without 
any mean-field or independent cluster approximation.

In this contribution, we propose to study the temperature and/or density 
dependence of the symmetry coefficient $c_{sym}$ in a lattice gas model.

\section{Description of the model}

We use a cubic lattice of linear site $L=20$, where each cell $i$ is 
characterised by four degrees of freedom : one discrete variable $\sigma_i$
for isospin ($\sigma_p=1, \sigma_n=-1, \sigma_0=0$) and three continuous 
variables $\vec{p_i}$ for the momentum.

The hamiltonian of the system follows:

\begin{equation}
H=\sum_{\langle i,j \rangle} \epsilon_{\sigma_i \sigma_j} \sigma_i \sigma_j
 + \sum_{\sigma_i= \sigma_j=1, (i \neq j)} \frac{I_c}{r_{ij}} + 
 \sum_{i=1}^{L^3} \frac{p_i^2}{2m} \sigma_i^2
\label{H}
\end{equation}

where $<i,j>$ are nearest neighbour cells, $\epsilon_{\sigma_i \sigma_j}$ is the 
coupling between nearest neighbour 
($\epsilon_{1}=0, \epsilon_{-1}=5.5 MeV$), $I_c(=1.44MeV/fm)$ is the Coulomb interaction 
between all protons in the lattice, and $r_{ij}$ is the distance between sites 
$i$ and $j$.
The last part of the interaction is the kinetic term where $m=939 MeV$ is the
nucleon mass.

Calculations are made in the isobar canonical ensemble, which has been shown to
 be the correct canonical ensemble to describe unbound systems in the vacuum
~\cite{FGAnnalsPhys}.  

The partition fonction is then:

\begin{equation}
Z = \sum_{(n)} exp \left(- \beta \left( H^{(n)} + P  R^{3(n)} \right)  \right)
\label{partition}
\end{equation}

where the sum runs over all the possible realisations of the system, and 
$R^{3(n)}$ is the global extension of the system for each 
partition $(n)$ define as:

\begin{equation}
R^{3(n)} = \frac{2 \left(\sum r_i^3 \sigma_i^2\right)^{(n)}}{\left(\sum{\sigma_i^2}\right)^{(n)}}
\label{equR}
\end{equation}

The statistical average at a given value of $T$ and $P$ ($\langle R^{3(n)} \rangle$) will be noted $R^3$ in the following.

\section{Phase transition}

In its original version with only one type of particles and closest neighbours 
interaction, the Lattice Gas Model (LGM) is well known to exhibit a first order 
transition line at low temperature ending to a second order critical point
~\cite{Campi,FGhdr}.
In this case, with a short range isovector coupling and a long range repulsive 
interaction, the phase diagram is not qualitatively modified~\cite{Lehaut}.
Indeed we find a phase coexistence line ended by the critical point, characterized by its temperature $T_c$ and pressure $P_c$.
In the following, all calculations are made at the subcritical pressure 
$P=5.\times 10^{-5}MeV.fm^{-3}$.

In table~\ref{tablett}, we compile the transition temperature $T_t$ for 
different systems at the chosen pressure.
The inclusion of Coulomb and isospin dependence results in a lower transition 
temperature than for a scalar interaction without coulomb.
With the realistic hamiltonian, in all cases the isospin dependence is weak on 
the transition temperature, less than few percents.
\begin{table}[!h]
  \begin{center}
    \begin{tabular}{r|c c c c c}
      System & $(75,75)_{scalar}$ & $(75,75)$ & $(85,65)$ & $(88,62)$ & $(91,59)$\\
      \hline
      $T_t (MeV)$ & 3.17 & 2.27 & 2.285 & 2.290 & 2.295 \\
    \end{tabular}
    \caption{Liquid-Gas transition temperature of different systems of $A=150$ nucleons at subcritical pressure $P=5.10^{-5}MeV.fm^{-3}$.}
    \label{tablett}
  \end{center}
\end{table}

To explore the finite temperature symmetry energy of the model, we will now use a macroscopic parametrisation in order to connect the microscopic and 
macroscopic properties of these systems.

\section{From microscopic to macroscopic}

As shown in the previous section, finite temperature Lattice Gas systems are  
strongly dishomogeneous and clusterized because of the presence of the phase 
transition.
In this situation, it is not clear whether the global energetics of the system,
including its symmetry properties, can be described by a macroscopic
parametrisation depending on the average density as in the nuclear mean-field.

To explore this issue and in order to extract the symmetry energy of our systems, we use a liquid-drop (macroscopic) parametrisation for the interaction energy of the system, which reads:
\begin{equation}
  E_{int}^{LD}(\delta,\rho,T)= a_v(\rho,T)A+a_s(\rho,T)A^{2/3}+c_{sym}(\rho,T)A\delta^2 + \alpha_c(\rho,T)\frac{Z^2}{R}
\label{LDParametrisation}
\end{equation}
Here $\delta$ is the isospin asymmetry $(N-Z)/(N+Z)$, $T$ is the temperature and
$\rho=A/(4/3\pi R^3)$ is an estimation of the average density of the system, 
where the mean cubic radius from eq.\ref{equR} is calculated, excluding the 
monomers ($A=1$).

This parametrisation uses four macroscopic parameters: $a_v$ is associated to 
the volume energy, $a_s$ corresponds to the surface energy, $c_{sym}$ is the 
coefficient related to the symmetry energy, $\alpha_c$ corresponds to the 
Coulomb contribution of this interaction.

\subsection{Temperature dependence}

\begin{figure}[!h]
  \includegraphics[width=1\columnwidth]{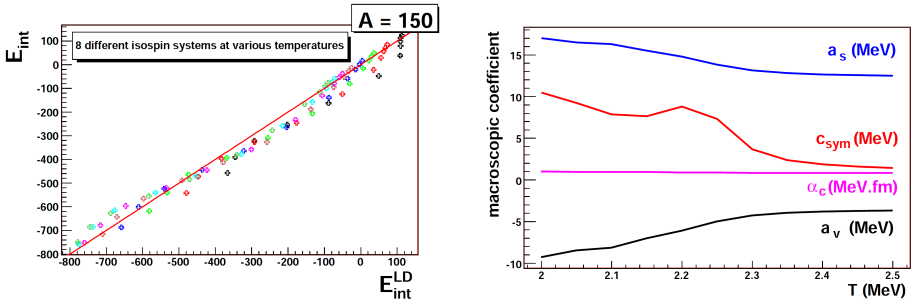}
  \caption{Left: Correlation between the liquid drop parametrisation (eq.\ref{LDParametrisation}) and the average LGM energy for eight different systems at different temperatures.
  Right: Evolution of the macroscopic coefficients with the temperature.}
  \label{validation} 
\end{figure}

In order to extract all these parameters, we fit our systems at 
different temperatures.
The results obtained from the parametrisation are plotted versus the true values
 coming out from the simulation on the figure~\ref{validation}\emph{-left}.
We observe a good agreement between the two results (close to the line 
$E_{int}^{LD}=E_{int}$).
The values of the macroscopic coefficients are plotted as a function of the 
temperature on the figure~\ref{validation}\emph{-right}. 
$\alpha_c$ is constant with temperature, while $a_v$ and $a_s$ decrease with 
increasing temperature, and
$c_{sym}$ decreases with increasing temperature with a bump around the 
transition temperature for the considered pressure ($T_t \approx 2.25 MeV$).

A very strong temperature dependence is observed, at variance with nuclear
mean-field calculations~\cite{Chen}.
However, up to now, we have extracted the temperature dependence of the 
macroscopic coefficient without looking at the density dependence, while 
a temperature variation induces a change in the average system size for these 
constant pressure calculations~\cite{samaddar}.

\subsection{Density dependence}

\begin{figure}[!h]
  \includegraphics[width=1\columnwidth]{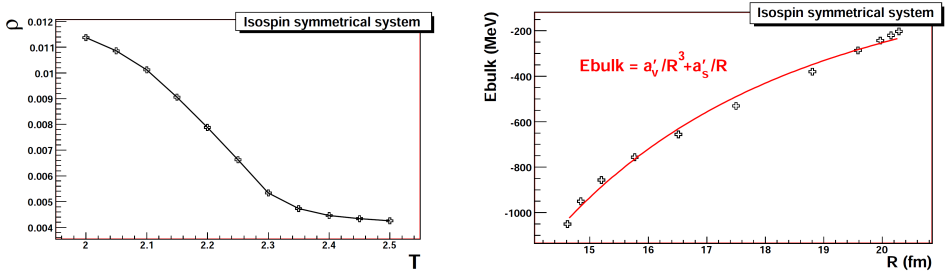}
  \caption{Left: density evolution in fonction of the temperature.
  Right: Total bulk energy in fonction of the radius of the system.}
  \label{density}
\end{figure}

On figure~\ref{density}\emph{-left}, we plot the evolution of the density with 
the temperature.
We observe a decrease of the density with increasing temperature similar to the 
behavior of $a_v$ and $a_s$.

The temperature variation of $a_v$ and $a_s$ can be easily understood as a 
simple effect of this density change, as we now explain.

To avoid the interference of the coulomb and symmetry effects, we consider an 
isospin symmetric system ($N=75$, $Z=75$) and switch off the coulomb 
interaction.

To distinguish the respective role of $\rho$ and $T$, we sort out the 
temperature dependence induced by the density variation $\rho(T)$ as:

\begin{equation}
  a_{v,s}(\rho,T) \propto f_{v,s}(\rho(T)) a'_{v,s}(\rho,T)
  \label{equdensdep}
\end{equation}

If the temperature dependence shown in figure~\ref{validation} is only due to 
the change with temperature of the average size as the mean-field approximation,
 eq.\ref{equdensdep} should be fulfilled with $a'_v$ and $a'_s$ as constants.
This is indeed what is observed once the total bulk energy is plotted as a 
function of the mean radius, as shown in figure~\ref{density}\emph{-right}. 
The observed functional dependence of $f(\rho)$ can be easily understood from 
simple geometrical considerations :

\begin{equation}
  \begin{array}{lr}
    f_v(\rho(T))\propto \frac{\rho(T)}{\rho_0} \propto R^{-3}(T) &
    f_s(\rho(T))\propto \frac{S(T)}{V(T)} \propto R^{-1}(T) \\
  \end{array}
\end{equation}

where $S$ and $V$ represent the average system surface and volume.

A similar analysis for the symmetry coefficient with the realistic hamiltonian including Coulomb is currently under progress.

\section*{Conclusions}

In order to connect the microscopic properties of the nuclear equation of state 
(symmetry energy and its dependence on temperature and/or density) and 
experimental fragment observables like isoscaling, we have used a lattice gas 
model with an isospin dependent short range and a long range coulomb 
interaction.
In this case, we observe a liquid-gas transition at lower temperature than the 
liquid gas transition with a purely nuclear isospin independent interaction.
The coupling of an isovector and coulomb interaction induces another first 
order transition, a kind of fission-fusion transition at lower temperature than 
the liquid-gas one.
We have also established a macroscopic parametrisation of the model,
with four parameters which are function of temperature and density 
($a_v$ , $a_s$ ,$c_{sym}$, $\alpha_c$ ).
For isospin symmetric systems, and as long as the coulomb interaction is 
ignored, the whole temperature dependence appears to be exhausted by the total
 average density dependence, even in the phase transition region where the 
system is highly dishomogeneous and clusterised.
This finding suggests that it may be possible to access the density dependence 
of $c_{sym}$ in multifragmentation experiments, as it is actively searched for 
in several recent experimental analysis~\cite{Shetty, Lefevre,Botvina}.
The connection between isoscaling observables and the symmetry energy for finite
 temperature within this model is currently in progress~\cite{Lehaut}.

%
\end{document}